\documentclass{article}

\usepackage[utf8]{inputenc}
\usepackage{geometry}
\usepackage{setspace}
\usepackage{titling}
\usepackage{booktabs}
\usepackage{bm}
\usepackage{amsfonts,amssymb,amsbsy,amsmath,amsthm}
\usepackage{multirow}
\usepackage{latexsym,dsfont}
\usepackage{gensymb}
\usepackage{bbm}
\usepackage{graphicx}
\usepackage{wrapfig}
\usepackage{lipsum}
\usepackage[bottom]{footmisc}
\usepackage{multicol}
\usepackage{rotating}
\usepackage{textcomp}
\usepackage[dvipsnames,table,xcdraw]{xcolor}
\usepackage{tcolorbox}
\usepackage{mathtools}
\mathtoolsset{showonlyrefs}
\usepackage{natbib}
\usepackage{hyperref}
\usepackage{authblk}
\usepackage{subfig}
\usepackage{sidecap}
\usepackage{fancyvrb}

\title{Spatial extremal modelling: A case study on the interplay between margins and dependence}
\author[1,*]{L. Kakampakou}
\author[2]{E.S. Simpson}
\author[1]{J.L. Wadsworth}

\affil[1]{\small 
School of Mathematical Sciences, Lancaster University, LA1 4YF, United Kingdom}
\affil[2]{Department of Statistical Science, University College London, WC1E 6BT, United Kingdom}
\affil[*]{Correspondence to:  \href{mailto:l.kakampakou1@lancaster.ac.uk}{l.kakampakou1@lancaster.ac.uk}}

\date{\today}

\begin{document}

\maketitle

\begin{abstract}
It is no secret that statistical modelling often involves making simplifying assumptions when attempting to study complex stochastic phenomena. Spatial modelling of extreme values is no exception, with one of the most common such assumptions being stationarity in the marginal and/or dependence features. If non-stationarity has been detected in the marginal distributions, it is tempting to try to model this while assuming stationarity in the dependence, without necessarily putting this latter assumption through thorough testing.
However, margins and dependence are often intricately connected and the detection of non-stationarity in one feature might affect the detection of non-stationarity in the other. This work is an in-depth case study of this interrelationship, with a particular focus on a spatio-temporal environmental application exhibiting well-documented marginal non-stationarity. Specifically, we compare and contrast four different marginal detrending approaches in terms of our post-detrending ability to detect temporal non-stationarity in the spatial extremal dependence structure of a sea surface temperature dataset from the Red Sea.
\end{abstract}

\noindent%
\vfill

\section{Introduction} \label{sec:intro}

In environmental applications it is not uncommon to observe non-stationary behaviour in the extreme values of a random variable. However, extreme value modelling often relies on a simplifying assumption of stationarity. In practice, non-stationarity may be present in different aspects of the data generating process; the more aspects it affects, the greater the degree of modelling complexity. Marginal non-stationarity refers to the situation where realisations of a univariate random variable are non-identically distributed. In environmental data, this varying behaviour typically depends on time, although other covariates, that themselves evolve with time, are normally the drivers of change. Various methods for modelling marginal non-stationarity of extremes have been well explored in the literature. One of the main approaches is to incorporate covariates directly into the parameters of univariate models for extreme value analysis. Typically this will be the generalised Pareto model for excesses of a high threshold: see for example \citet{Davison1990}, \citet{chavezdemoulin2005} and \citet{Youngman2019}. Section \ref{sec:marginal} contains further details on these and other methods for modelling non-stationarity. 
 
Non-stationarity may also be encountered in the extremal dependence structure of a random vector. That is, the extremal dependence between two or more components of a random vector may change with covariates. In such instances, it is common in analyses of multivariate or spatial extremes to handle marginal and dependence modelling as a two-step procedure, implementing marginal detrending methods first and then applying non-stationary methods for dependence modelling to the marginally detrended data. 
Although alternative methods that model marginal and copula behaviour jointly in a single step exist, they are not as numerous and their application in extreme value modelling is scarce in practice. One reason for this is that univariate extreme value distributions are suited to marginal upper tails, while multivariate extremal dependence models rely on potentially different upper tail definitions. A mismatch between marginal and copula support impedes joint modelling considerations. When the multivariate definition of extremity follows as a direct generalisation of the univariate one, as in the case of modelling maxima, \citet{Padoan2010} and \citet{Ribatet2012} develop such joint modelling procedures for max-stable processes, in a frequentist and Bayesian realm respectively. In a similar manner, multivariate generalised Pareto distributions \citep{ROOTZEN2006,ROOTZEN2018b} allow for joint marginal-dependence modelling, which could be extended to cover non-stationarity, while similar ideas could potentially be applied to spatial generalised Pareto processes \citep{Buishand2008,Ferreira2014}. However, both max-stable and generalised Pareto processes yield very restrictive dependence structures and are therefore unlikely to capture the complex dependence features often present in environmental applications \citep{Huser2022}. Even putting these issues aside, pursuing joint marginal-copula modelling brings about another important risk, even in the stationary case, that misspecification in one form --- the margins or dependence --- can affect the other. Such spillover is likely worsened when non-stationarity is present. In this paper, we adopt the more established two-stage approach, thus making the task of disentangling the relationship between margins and dependence, which is the focus of this work, as clear and straightforward as possible.

Available methods for non-stationary dependence modelling include the following. In a bivariate setting, \citet{deCarvalho2014} develop a semi-parametric model which incorporates covariates into the modelling of the so-called \textit{spectral density}, thus capturing the effect of explanatory variables on joint extremes. In a multivariate setting, \citet{Castro-Camilo2018} and \citet{Mhalla2019regression} also focus on spectral density functions with the former developing non-parametric regression-based methods and the latter building a vector generalised additive model to capture covariate-varying extremal dependence. \citet{Lee2024} model covariate(time)-varying extremal dependence in various summary statistics. \citet{Murphy-Barltrop2022} model temporally varying extremal dependence in a bivariate setting via non-stationary extensions of the so-called \textit{angular dependence function} and \textit{return curves}, while \citet{Mhalla2019} also make use of non-stationary angular dependence functions to account for covariate influence. 

In a spatial setting, modelling changes in extremal dependence using space-related covariates, which may be expected when the spatial domain is large or morphologically diverse, has received the most attention; see for example \citet{Huser2016}, \citet{Blanchet2011}, \citet{Richards2021}. Much less attention has been dedicated, however, to time-related covariates which, given the ongoing climate crisis, may have considerable effects on the spatial extremal dependence structure over time, especially in applications spanning a large temporal domain. 
A noteworthy exception of research in this direction is that of \citet{Healy2023} who model temporal non-stationarity in both marginal and dependence features. They combine information on Irish temperatures from both observational and climate-model sources and consider time-varying dependence via the framework of $r$-Pareto processes \citep{dombry2015}.

Of the aforementioned examples, \citet{Healy2023} is the only one dealing with spatial aspects of time-varying extremal dependence; the remaining do so in a multivariate setting. To the best of our knowledge, there is no other contribution where this problem is tackled in a fully spatial context available in the literature. What is more, \citet{Healy2023} focus on a class of models that can only accommodate a particular type of dependence, namely asymptotic dependence (see Section \ref{sec:dep measures} for details) which, although appropriate for their application, is often unrealistic for similar spatial environmental datasets.

It is this lack of available literature that provided the initial motivation for the present work. Our study is focused on a dataset with well documented marginal non-stationarities \citep{Huser2016}, which was also believed to showcase non-stationary behaviour in its spatial extremal dependence structure. The data comprise daily sea surface temperature values from the Gulf of Suez spanning across $31$ years ($1985$ to $2015$ inclusive). As is common in analyses of extremes, we focused our attention on the summer data (July-August-September) from that period, to reduce the degree of seasonality to be modelled, while considering the period of highest temperatures. The original data are available for the whole of the Red Sea, on a daily basis and a very fine resolution grid of $0.05\degree \times 0.05\degree$ \citep{Donlon2012}.  Initial analysis of spatial dependence changes over time for this dataset suggested the need for adapting spatial dependence models to account for time-varying dependence (see top left plot of Figure \ref{fig:chi diff}). 

However, following our initial exploratory analysis, we tested various approaches to modelling marginal non-stationarity, finding that subsequent results concerning changes in spatial extremal dependence were highly sensitive to the marginal model used. We thus came to the conclusion that the deciding factor behind whether or not temporal non-stationarity is detectable in the spatial extremal dependence structure of the data critically relies on the marginal detrending approach pursued and that any resultant analysis is, in our case, extremely sensitive to marginal modelling choices. As such, instead of focusing on adapting existing models to accommodate changes in spatial dependence over time, the aim of this paper is to serve as a cautionary tale of the misleading effects the marginal and dependence structure interrelationship can have on analyses of extremes. Our main objective is to investigate the effects of four different marginal detrending approaches on spatial extremal dependence. In particular, we use graphical tools to assess whether the apparent spatial extremal dependence changes over time under each marginal detrending procedure. We note here that the dataset in question displays strong autocorrelation; such a feature will remain after modelling trends in the data, but can affect visual assessment of stationarity. Graphical tools should therefore be interpreted with some degree of caution. We further assess the impact of each marginal model on the spatial extremal dependence under a (potentially false) assumption of temporal stationarity by applying the conditional spatial extremes model \citep{Wadsworth2022} to all four marginally treated versions of the Gulf of Suez dataset and comparing model‐based dependence features. This particular spatial model was selected because of its computational and modelling flexibility, which allow for increased scalability (in terms of the number of spatial locations that can be handled) compared to most alternative models available, as well as simpler characterisation/modelling of different types of extremal dependence. 

The remainder of the paper is outlined as follows. Section \ref{sec:background} provides background information for our statistical approach. Section \ref{sec:methods} describes the four marginal detrending methods employed. Section \ref{sec:results} is dedicated to the analysis of the Gulf of Suez data. Finally, we conclude in Section \ref{sec:discussion} with a short discussion. 

\section{Statistical background} \label{sec:background}

\subsection{Marginal modelling of extremes} \label{sec:marginal}
Marginal modelling in extremes concerns the study of univariate extreme values which appear either in isolation, or as components of a higher dimensional random vector. When studying a random variable $X$, asymptotic theory dictates that the distribution of threshold exceedances $X-u\mid X>u$ of a sufficiently high threshold $u$ can usually be approximated by the limiting form $G$ in \eqref{GPD}, known as the generalized Pareto distribution (GPD):
\begin{equation} \label{GPD}
G(x)=     
\begin{cases}
    1-\left(1+\xi x/\sigma_u \right)_{+}^{-1/\xi}, & \text{for } \xi \neq 0, \sigma_u >0, x > 0, \\
    1-\exp\left(-x/\sigma_u \right), & \text{for } \xi \to 0, \sigma_u>0, x>0,
\end{cases}    
\end{equation}
for $a_+  =\max(a,0)$; we write $X-u\mid X>u \sim \text{GPD}(\sigma_u, \xi)$. The GPD model is suitable for capturing the tail behaviour of stationary processes. When marginal non-stationarity is present, \citet{Davison1990} proposed incorporating covariates into the parameters of the GPD. That is, given a marginally non-stationary process $\{X_t\}$ with associated covariates $\mathbf{Z}_t$, one can model $\left(X_t-u\mid X_t>u, \mathbf{Z}_t=\mathbf{z}_t\right) \sim \text{GPD}\left(\sigma_u(\mathbf{z}_t), \xi(\mathbf{z}_t)\right)$ for a high enough threshold $u$. Note that covariate modelling here is performed according to the principles of generalised linear models, where typically we set $\log \sigma_u(\mathbf{z}_t) = \boldsymbol{\beta}_{\sigma}^T\mathbf{z}_t$, $\xi(\mathbf{z}_t) = \boldsymbol{\beta}_{\xi}^T \mathbf{z}_t$. Subsequent work has considered more flexible covariate formulations, such as those provided by generalised additive models (GAMs); see, for example, \citet{chavezdemoulin2005} and \citet{Youngman2019}.

One problem with such techniques is that their definition of extremity often depends on a constant threshold, but under non-stationarity, the extremes are suitably defined through the corresponding covariate levels as well. To address this issue, \citet{Northrop2011} and \citet{Jonathan2013} define covariate-varying thresholds. Another concern is that these techniques focus on modelling non-stationary behaviour only in the tail of a process. However, it is often of interest to model non-stationarity in the body of the data as well, since extremes across components of a random vector do not necessarily occur simultaneously. That is, we may observe extremity in one component but moderate levels of another. If marginal non-stationarity in the body of the distribution is neglected, then we cannot model this dependence appropriately.

To overcome these issues, \citet{Eastoe2009} borrow techniques from time-series modelling. They introduce a two-step procedure comprising a pre-processing step, aiming to remove most of the non-stationarity present in the body of the data, and a non-stationary GPD step, to remove any leftover trends in the tails. The pre-processing step consists of applying a Box-Cox location-scale transformation of the form 
\begin{equation}\label{eastoe detrend}
   \frac{X_t^{\lambda(\mathbf{z}_t)}-1}{\lambda(\mathbf{z}_t)} = \mu(\mathbf{z}_t) + \sigma(\mathbf{z}_t)R_t
\end{equation} 
to the marginally non-stationary process $\{X_t\}$, where $\{R_t\}$ is approximately stationary and $\lambda$, $\mu$ and $\log(\sigma)$ are linear functions of covariates $\mathbf{z}_t$. 
The process $\{R_t\}$ is assigned a Gaussian distribution, so that specification \eqref{eastoe detrend} has a likelihood that is maximised to provide parameter estimates. This is then followed by a non-stationary GPD fit to the residual process $\{R_t\}$, using the methodology of \citet{Davison1990}. If any trends are identified in the scale and shape parameters, $\sigma_u(\mathbf{z}_t)$ and $\xi(\mathbf{z}_t)$ respectively, they are removed via the probability integral transformation, using a semi-parametric approach to estimation of the distribution of $R_t$ \citep{TawnColesSemipar}. This consists of a rank transform for data below a high marginal threshold, $u$, which are now assumed stationary, and the GPD cumulative distribution function for values above $u$. To make things more explicit, let $r_1, \ldots, r_n$ be observations from the process $\{R_t\}$ at times $1, \ldots, n$, and define
\begin{equation}
    \hat{F}_{R_{t}}(r\mid\mathbf{Z}_t=\mathbf{z}_t)=
    \begin{cases}
        \Sigma_{t=1}^n \mathbbm{1}(r_{t}\leq r)/(n+1), & \qquad r\leq u \\
        1-(1-u)\left[1+\hat{\xi}(\mathbf{z}_t)(r-u)/\hat{\sigma}_u(\mathbf{z}_t)\right]^{-1/\hat{\xi}(\mathbf{z}_t)}_+, & \qquad r>u, \\
    \end{cases}
\label{semipar pit}    
\end{equation}
where $\hat{\sigma}_u(\mathbf{z}_t)$ and $\hat{\xi}(\mathbf{z}_t)$ are the  maximum likelihood estimates of the scale and shape parameters respectively, which are estimated by the non-stationary GPD step. Applying $\hat{F}_{R_{t}}$ to the marginal process $\{R_{t}\}$ achieves standardisation to a standard uniform scale. 
Finally, we note that some more recent work replaces \citet{Eastoe2009}'s pre-processing step with a more elaborate one which, instead of assuming linear parametric forms for the covariate functions, makes use of the GAM framework to allow parameters $\mu(\mathbf{z}_t), \sigma(\mathbf{z}_t)$ of \eqref{eastoe detrend} and $\sigma_u, \xi$ of the non-stationary version of \eqref{GPD} to vary smoothly with covariates \citep{Murphy-Barltrop2022}. This extra flexibility is desirable when the observed non-stationarities are too complicated to be adequately captured by parametric forms.

\subsection{Measures of extremal dependence} \label{sec:dep measures}
A key endeavour of many multivariate analyses is to characterise the extremal dependence between $D\geq2$ random variables. Apart from the obvious exploratory benefits of understanding the data better by uncovering its dependence structure before attempting any modelling, another reason why such a task is of importance is that many of the available models --- and their underpinning assumptions --- can only accommodate data that fall under a single category of dependence classes. 
Hence, $D$-variate measures of extremal dependence are often examined for dependence-class assessment and categorisation in multivariate analyses.
In the spatial case however, one can often assume that bivariate distributions ($D=2$) of the spatial process at location pairs are very informative and, thus, looking at bivariate versions of extremal dependence measures for different pairs is all one needs to characterise the extremal dependence of the process at given distances. 

A widely used dependence measure in practice is the so-called \textit{tail dependence coefficient}, typically denoted by $\chi$. Let $\{Y(\boldsymbol{s}): \boldsymbol{s} \in S \subset \mathbb{R}^2 \}$ be a stationary and isotropic spatial random field and $\left(Y(\boldsymbol{s}_1),\ldots,Y(\boldsymbol{s}_D)\right)$ the $D$-dimensional random vector comprised of the realisations of $\{Y(\boldsymbol{s})\}$ at the observed spatial locations $\boldsymbol{s}_i$, $i=1, \ldots, D$. Letting $Y(\boldsymbol{s}_i)\sim F$ denote the common marginal distribution, the coefficient of tail dependence $\chi(h_{jk})$ is given by $\chi(h_{jk}) = \lim_{u \to 1} \chi_u(h_{jk})$, with
\begin{equation}\label{spat chi}
   \chi_u(h_{jk}) = \Pr \left\{F(Y(\boldsymbol{s}_j))>u \mid F(Y(\boldsymbol{s}_k))>u\right\}, \quad u\in[0,1],
\end{equation}
where $h_{jk}=\|\boldsymbol{s}_j - \boldsymbol{s}_k \|$ is the distance between sites $j$ and $k$ \big{(}$j\neq k$, $j,k\in\{1,\ldots,D\}$\big{)}; we will take this to be the Euclidean distance for the remainder of the paper. The quantity $\chi(h_{jk})$ is the limiting probability of the process being extreme at location $\boldsymbol{s}_j$ given that it is also extreme at location $\boldsymbol{s}_k$ and characterises the extremal dependence between the pair $(j,k)$ at distance $h_{jk}$ apart. A process is said to be \textit{asymptotically dependent} (AD) in the spatial domain $S$ if $\chi(h_{jk})>0$ for all $h_{jk}$, $j,k\in\{1,\ldots, D\}$. In practical terms, asymptotic dependence (AD) means that the variables, or realisations of the spatial process at different locations, can take their most extreme values simultaneously. Conversely, the case where $\chi(h_{jk})=0$ for all $h_{jk}$, $j,k\in\{1,\ldots, D\}$, defines an \textit{asymptotically independent} (AI) process in the spatial domain $S$, which means that the spatial extent of the extremes becomes more and more localised as the level of extremity increases. In such cases, the measure $\chi(h_{jk})$ does not give us any additional information on the joint tail dependence. A complementary statistic to $\chi(h_{jk})$ that is more informative under asymptotic independence (AI) is the \textit{residual tail dependence coefficient}, $\eta(h_{jk}) \in (0,1]$, of \citet{Ledford1996}, which may be defined through the relation 
\begin{equation}\label{eta}
    \Pr\left\{F(Y(\boldsymbol{s}_j))>u,F(Y(\boldsymbol{s}_k))>u\right\}= \mathcal{L}(1-u)(1-u)^{1/\eta(h_{jk})}, \quad u \to 1,
\end{equation}
where $\mathcal{L}$ is a \textit{slowly varying function} such that $\mathcal{L}(cr)/\mathcal{L}(r)\to 1$ as $r\to 0$ for all positive constants $c$. Note that relation \eqref{eta} implies $\eta(h_{jk}) = \lim_{u\to 1}\eta_u(h_{jk})$, with
\begin{equation}\label{spat eta}
    \eta_u(h_{jk})=\frac{\log(1-u)}{\log\left(\Pr\left\{ F(Y(\boldsymbol{s}_j))>u,F(Y(\boldsymbol{s}_k))>u\right\}\right)}, \quad u\in[0,1],
\end{equation}
$j\neq k$, $j,k\in\{1,\ldots,D\}$. 
The boundary case $\eta(h_{jk})=1$ corresponds to $\chi(h_{jk}) = \lim_{u\to 1}\mathcal{L}(1-u)$. Hence, when $\eta(h_{jk})=1$ for all $h_{jk}$, $j,k\in\{1, \ldots, D\}$, and $\mathcal{L}(1-u) \not\to 0$ as $u\to 1$ we get $\chi(h_{jk})>0$ for all $h_{jk}$, $j,k\in\{1, \ldots, D\}$; that is, the process is AD in the spatial domain $S$. The more interesting case where either $\eta(h_{jk})<1$ for all $h_{jk}$, $j,k\in\{1, \ldots, D\}$, or $\eta(h_{jk})=1$ for all $h_{jk}$, $j,k\in\{1, \ldots, D\}$, and $\mathcal{L}(1-u) \to 0$ as $u\to 1$, leads to $\chi(h_{jk})=0$ for all $h_{jk}$, $j,k\in\{1, \ldots, D\}$, and signifies that the process is AI in $S$, which can be further categorised into \textit{negative extremal association} when $\eta(h_{jk}) \in (0,1/2)$, \textit{near independence} when $\eta(h_{jk})=1/2$ and \textit{positive extremal association} when $\eta(h_{jk})\in(1/2,1]$. The latter case is of particular interest to environmental applications which usually exhibit positive dependence over space. Because we cannot typically estimate the limiting quantities $(\chi(h_{jk}),\eta(h_{jk}))$,  the behaviour of $(\chi_u(h_{jk}),\eta_u(h_{jk}))$ as $u \to 1$ is used instead to help inform us about the extremal dependence structure of the process.

As already mentioned, distinguishing between extremal dependence classes is important for modelling. Doing so in practice, however, is a non-trivial task. With spatial data, for example, we might have $\chi_u(h_{jk})>0$ at all observable high quantiles, $u$, and pairs of sites, but $\chi_u \searrow 0$ as $u \nearrow 1$. It is frequently encountered in practice that the dependence of environmental processes weakens as events become more extreme with the very severe extreme events becoming more spatially localised \citep{Huser2022}. 
One might also observe that the dependence class of the process is not the same for all distances, but changes from AD to AI after a distance $\Delta$, or with respect to other covariates.
This creates additional modelling difficulties and highlights the importance of thorough exploratory investigation of the dependence structure as a first step to overcoming said difficulties.

\subsection{Conditional spatial extremes model} \label{sec:cse}

So-called \textit{conditional extremes methods} for extremal dependence modelling were first developed in a multivariate setting by \citet{Heffernan2004} and further studied by \citet{Heffernan2007}. In contrast to previously existing methods that relied on the limiting behaviour of a $D$-dimensional random vector when all of its components become simultaneously extreme at the same rate, the authors developed novel methodology based on a limiting assumption on the joint tail of a vector given that one of its components is extreme. This alternate limiting assumption of the \citet{Heffernan2004} model is much less restrictive, allowing for increased flexibility both in dependence modelling and in fitting scalability. With these benefits in mind, \citet{Wadsworth2022} extended the work of \citet{Heffernan2004} and \citet{Heffernan2007} to a spatial setting, termed the conditional spatial extremes (CSE) model. 

Let $\{Y(\boldsymbol{s}): \boldsymbol{s} \in S \subset \mathbb{R}^2 \}$ be a stationary and isotropic spatial random field with exponential-tailed margins, i.e.\ $\Pr(Y(\boldsymbol{s}_j)>x) \sim ce^{-x}$, $c>0$, as $x\to\infty$; a typical marginal choice is standard Laplace, which we also adopt in our subsequent analyses.
For a high marginal threshold $t$, \citet{Wadsworth2022} show that for many dependence structures, a reasonable assumption is
\begin{equation}\label{cse model}
    \left\{Y(\boldsymbol{s}) \mid Y(\boldsymbol{s}_0)>t: \boldsymbol{s} \in S\right\} \stackrel{\text{d}}{\approx} \left\{a_{\boldsymbol{s}-\boldsymbol{s}_0}\left(Y(\boldsymbol{s}_0)\right) + b_{\boldsymbol{s}-\boldsymbol{s}_0}\left(Y(\boldsymbol{s}_0)\right)Z^0(\boldsymbol{s}): \boldsymbol{s}\in S\right\},
\end{equation}
which can be used to model the behaviour of the process $\{Y(\boldsymbol{s})\}$ given an extreme value has occurred at some location $\boldsymbol{s}_0$. Specifying functional forms for $a_{\boldsymbol{s}-\boldsymbol{s}_0}$ and $b_{\boldsymbol{s}-\boldsymbol{s}_0}$, where $a_{\boldsymbol{s}-\boldsymbol{s}_0}: \mathbb{R} \rightarrow \mathbb{R}$, $a_0(y)=y$, and $b_{\boldsymbol{s}-\boldsymbol{s}_0}: \mathbb{R} \rightarrow (0, \infty)$, $\boldsymbol{s} \in S$, and a distributional form for the so-called \textit{residual process} $\{Z^0\}$ --- satisfying the condition $Z^0(\boldsymbol{s}_0)=0$ --- completes the model. The dependence structure of the residual process $\{Z^0\}$ is assumed to be Gaussian in nature, while its margins are modelled via the so-called \textit{delta-Laplace} distribution. Inference for the model in \eqref{cse model} is performed based on a composite likelihood scheme, combining likelihood contributions stemming from different conditioning locations (all or a subset of available locations can be selected), thus overcoming the problem of choice of conditioning location which is often not obvious in applications. Additional details on the CSE model are given in Appendix \ref{sec:appendix_cse}.

\section{Marginal detrending methods} \label{sec:methods}
In Sections \ref{sec:marg a} -- \ref{sec:marg d} we outline four approaches considered for capturing marginal non-stationarities in our dataset. In each case, our objective is to transform the data to follow a stationary standard Laplace distribution, i.e.\ with distribution function  
\begin{equation}
    F_L(x) = 
    \begin{cases}
    0.5 \exp\{x\}, &x \leq 0, \\
    1-0.5 \exp\{-x\}, &x > 0, 
    \end{cases}
\end{equation}
enabling subsequent application of the CSE model. We do this in two steps. We first transform our data to the standard uniform scale by implementing each detrending procedure outlined below. Then, assuming all trends have been successfully removed and that, therefore, our data are now stationary, we apply $F_L^{-1}$, the Laplace inverse cumulative distribution function, to the uniform data to bring them to the required marginal scale. Figure \ref{fig:ts laplace} displays the data on Laplace margins following each of the four processing techniques. As noted in Section \ref{sec:intro}, strong temporal dependence between observations within a year can hinder visual assessment of stationarity from such plots. All marginal detrending procedures are implemented site-wise.

\subsection{Margins A} \label{sec:marg a}
\begin{figure}[t]
  \centering
  \includegraphics[width=0.48\textwidth]{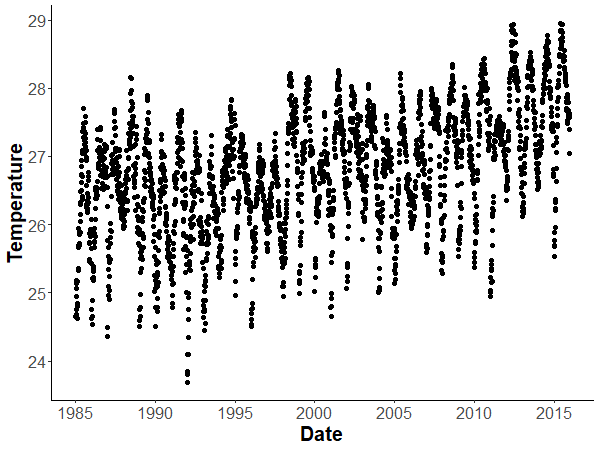}
  \caption{Time series plot at location $\boldsymbol{s}_{35}$ on the original marginal scale.}
  \label{fig:time series}
\end{figure}

The first marginal detrending approach considered comes from the simple but crude assumption that data are stationary within a year --- facilitated by considering data only from a single season --- but that their distribution may change over the years.
It consists of a simple rank transform applied in yearly blocks, whereby data corresponding to the same year are grouped, rank-transformed together, and collated post-transformation. Implementing this transformation separately for each year allows for long term and inter-annual trends --- which, according to Figure \ref{fig:time series}, both seem to be present for the Gulf of Suez dataset --- without the need to explicitly specify the form of such trends.

The main merit of such an approach lies precisely in its simplicity, which in turn leads to very fast and straightforward implementation. On the other hand, performing the rank operation yearly could lead to cruder empirical approximations of the underlying marginal distributions --- given only 92 data points per year are available in our case --- and introduces the repeated values that can be seen in the top-left panel of Figure \ref{fig:ts laplace} across the full, transformed time series.

\subsection{Margins B} \label{sec:marg b}
The \citet{Eastoe2009} approach has been widely used in applications to achieve data detrending at a marginal level (e.g.\ \citet{Winter2016}, \citet{Winter2017}). We employ this, in particular its more flexible GAM extension, as our second detrending technique. As evidenced in Figure \ref{fig:time series}, the Gulf of Suez dataset is characterised by strong temporal trends of both cyclical (short-term) and linear (long-term) nature. We endeavour to remove such trends site-wise, via the GAM modelling framework of \citet{wood2017generalized}, as implemented in \citet{Murphy-Barltrop2022}. 

\begin{figure}[t]
  \flushleft
  \includegraphics[width=0.48\textwidth]{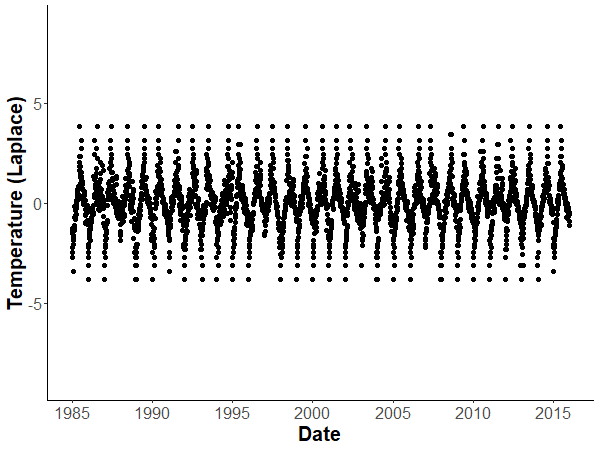}
  \hfill
  \includegraphics[width=0.48\textwidth]{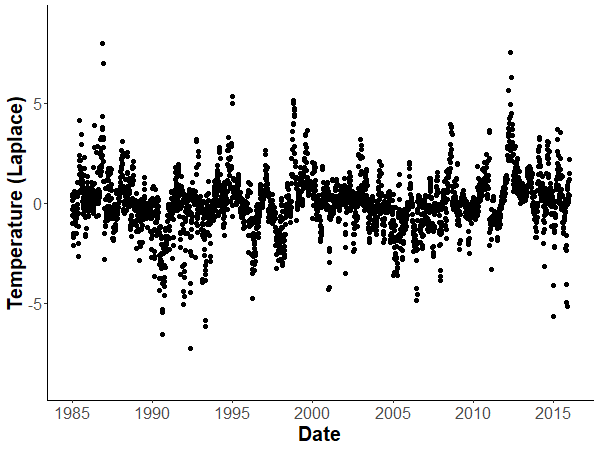}
  \hfill
  \includegraphics[width=0.48\textwidth]{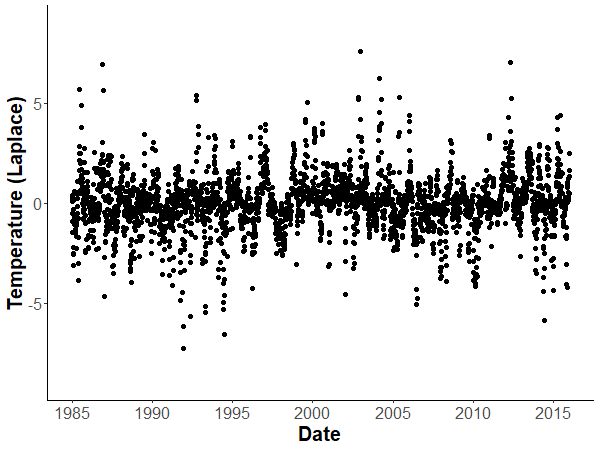}
  \hfill
  \includegraphics[width=0.48\textwidth]{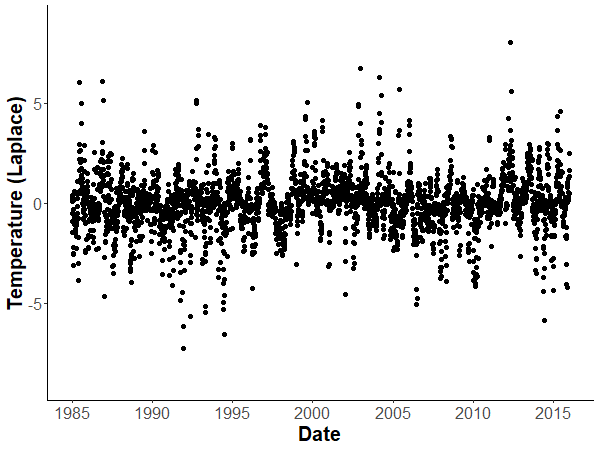}
  \caption{Time series for location $\boldsymbol{s}_{35}$ in standard Laplace margins. [Left to right and top to bottom:] A-B-C-D margins. Detrending is performed based on a $90\%$ constant marginal threshold for margins A-C, while for margins D an $83\%$ threshold was automatically selected for this location.}
  \label{fig:ts laplace}
\end{figure}

To be precise, let us denote the temperature process by $\{Y_t(\boldsymbol{s})\}$, $t=1,\ldots,T$. We are interested in applying detrending methods to all marginal time series $Y_{i,t} = Y_t(\boldsymbol{s}_i)$, for all $i\in\{1,\ldots,D\}$. For every $Y_{i,t}$, we allow the location, $\mu_i$, and scale, $\sigma_i$, parameters of \eqref{eastoe detrend} to vary smoothly with covariates $\mathbf{z}_t=\{1,\mbox{gmt}_t,d_t\}$, where $\mbox{gmt}_t$ denotes global mean temperature anomalies\footnote{Data form part of the HadCRUT5 dataset \citep{morice2021updated}.} at time $t$ and $d_t=\{1,\ldots,92\}$ denotes the day index of the process at time $t$. The reason $\mbox{gmt}_t$ is chosen instead of $t$ is that we found it explained more of the observed trends than simply $t$. Moreover, its use has been advocated by others (e.g.\ \citet{darcy2022}, \citet{Healy2023}) for better capturing long-term trends in environmental applications. Note that we do not include a Box-Cox parameter in this procedure for simplicity, as there is no obvious need for a changing shape parameter. Aiming to capture the long-term linear trend, a thin plate regression spline is fitted for the covariate $\mbox{gmt}_t$, while a cubic regression spline of dimension $92$ --- found to result in the most stationary-looking data amongst a number of different choices examined --- is used to capture the seasonal trends through the covariate $d_t$. The $\mu_i$ and $\sigma_i$ parameters are estimated by means of the R package \texttt{mgcv} \citep{wood2003,wood2011} and subsequently removed, resulting in residual time series $R_{i,t} = (Y_{i,t}-\mu_i(\mathbf{z}_t))/\sigma_i(\mathbf{z}_t)$ for all $i$. We then proceed by applying a non-stationary GPD to $R_{i,t}$ using a constant $90\%$ marginal threshold, thus aiming to remove any remaining trends in the tails. A likelihood ratio test is used to determine whether no, linear, seasonal or both trends should be added to the scale parameter of the GPD for each site $\boldsymbol{s}_i$. The shape parameter in all margins is assumed to be invariant to temporal trends. The detrended data are visualised on the standard Laplace scale in the top right plot of Figure \ref{fig:ts laplace}. 

\subsection{Margins C} \label{sec:marg c}
The third marginal detrending method examined is motivated by the unsatisfactory degree of detrending achieved in margins B. In particular, there seems to be an inconsistency in the number of exceedances of a fixed high threshold observed per year, with some years experiencing multiple exceedances and some very few, if any. It appears therefore that the detrending method described in Section \ref{sec:marg b} fails to capture the full structural complexity of the margins, resulting in leftover marginal ``irregularities" manifesting as inter-year variability. \citet{Eastoe2019} and \citet{Clarkson2023} have reported similar findings in their analyses of a river flow and a U.S. temperature dataset, respectively. To account for leftover inter-year variability they develop a random-effects-based methodology. Inspired by their method, we introduce an additional intermediate step to the aforementioned margins B procedure, acting as a crude approximation of the \citet{Eastoe2019} approach. 

Let $\mathcal{Y}_k$ denote the set of time points $t$ corresponding to year $k$, where $k=\{1, \ldots,31\}$. Since each year is comprised of $92$ summer-month observations, $\mathcal{Y}_1 = \{1, \ldots, 92\}$, $\mathcal{Y}_2 = \{93, \ldots, 184\}$, etc. The residual process $\{R_{i,t}\}$ is then adapted as follows:
\begin{equation}\label{intermediate step C}
    \frac{\{R_{i,t}\}_{t\in \mathcal{Y}_k} - m_k}{s_k},
\end{equation}
where $m_k$ is the yearly mean and $s_k$ is the yearly standard deviation. Both quantities are estimated over all locations $i \in \{1, \ldots, D\}$ and using all data points corresponding to time $t\in \mathcal{Y}_k$. The transformation in \eqref{intermediate step C} is applied to each $i\in \{1,\ldots, D\}$. This intermediate step is then followed by the final GPD step, for which we introduce a high but constant $90\%$ marginal threshold, similar to our approach for margins B.

\subsection{Margins D} \label{sec:marg d}
Our final detrending scheme builds upon the methodology for margins C, replacing the constant high marginal threshold feature with the automated threshold selection routine of \citet{murphy2023automated}, thus allowing for a different marginal threshold for each spatial location $\boldsymbol{s}_i$, $i \in \{1, \ldots, D\}$. 

The \citet{murphy2023automated} method is aimed at addressing the bias-variance trade-off that is inherent to the problem of threshold choice for GPD model fitting; namely that too low a threshold is likely to violate the assumptions upon which asymptotic justification for the model is based --- thus introducing bias in the GPD model fit --- while too high a threshold can lead to fewer excesses being used to fit the GPD model --- hence contributing to higher parameter uncertainty. The method focuses on selecting a constant threshold for independent and identically distributed univariate time series. Threshold selection is performed based on an algorithm that minimises the so-called \textit{expected quantile discrepancy} between the sample quantiles and the fitted GPD model quantiles. For more details, we refer the interested reader to the original paper.

We use code provided as supplementary material to the \citet{murphy2023automated} paper to apply their method to the pre-whitened and yearly adjusted series in \eqref{intermediate step C} used in margins C. A histogram of the resulting thresholds across all $D$ locations is given in Appendix \ref{sec:appendix_margD}. These thresholds are subsequently used for the implementation of the final GPD step, in accordance with all the aforementioned marginal procedures.
We note that neither the independence nor the identical distribution assumptions of the \citet{murphy2023automated} method are likely to hold in our case. The former is violated by the presence of temporal dependence mentioned in Section \ref{sec:intro}, while in the case of the latter, we allow for its possible violation, should the likelihood ratio test --- performed as part of the GPD fitting step --- deem a non-stationary fit appropriate. However, given the dimensionality of our data and the fact that few, if any, alternative automated approaches for appropriate threshold selection exist, we treat margins D and their subsequent analysis as a sensitivity experiment of the effect threshold choice may have on marginal features.
Visualisation of the resulting time series in standard Laplace margins (bottom right plot of Figure \ref{fig:ts laplace}) suggests potentially little effect of threshold sensitivity given the similarity of the series to the respective one for margins C (bottom left plot of Figure \ref{fig:ts laplace}).

\section{Spatial analysis of the Gulf of Suez dataset} \label{sec:results}
The Red Sea temperature dataset introduced in Section \ref{sec:intro} has been analysed in numerous other studies including \citet{Huser2016}, \citet{SIMPSON2021redsea} and \citet{Simpson2023}. In agreement with those studies, we found in exploratory analysis that the whole dataset showcases non-trivial non-stationarity in its spatial extremal dependence across space, with the Gulf of Suez having a slightly different behaviour to the north part of the Red Sea (latitude $>21.3\degree$), which in turn is significantly different to the south part of the Red Sea (latitude $<21.3\degree$). Hence the decision to focus on the Gulf of Suez where the assumption of stationarity in spatial extremal dependence over space seems reasonable. This smaller dataset now comprises a total of $D=510$ spatial locations, for which marginal time series consisting of $T=2852$ data points ($92$ per year) are available.
\begin{figure}[t]
  \flushleft
  \includegraphics[width=0.48\textwidth]{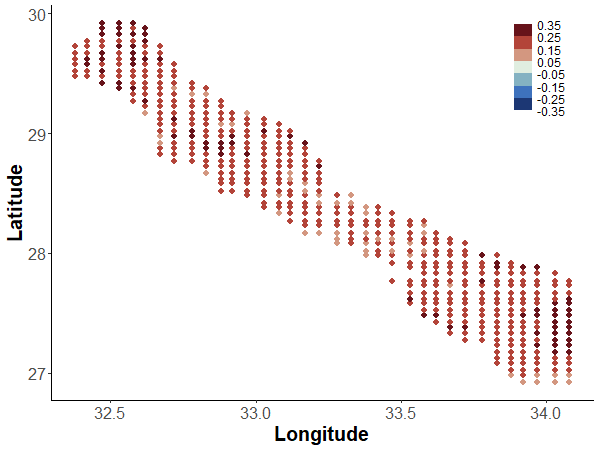}
  \hfill
  \includegraphics[width=0.48\textwidth]{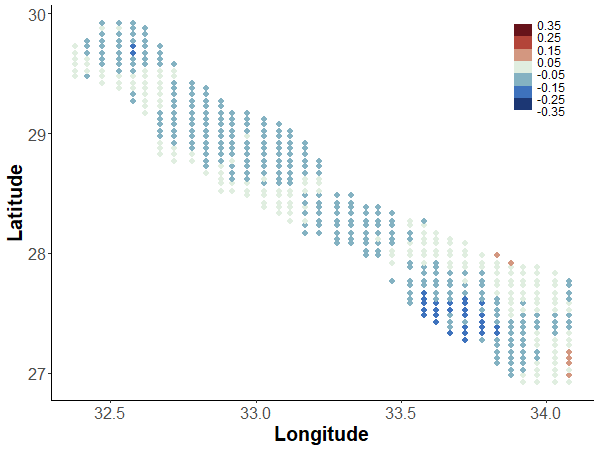}
  \hfill
  \includegraphics[width=0.48\textwidth]{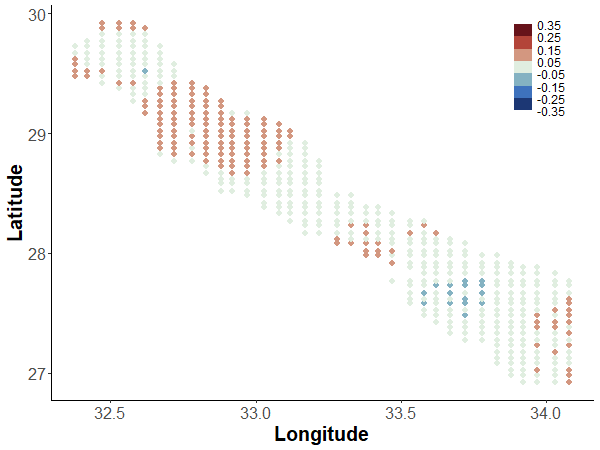}
  \hfill
  \includegraphics[width=0.48\textwidth]{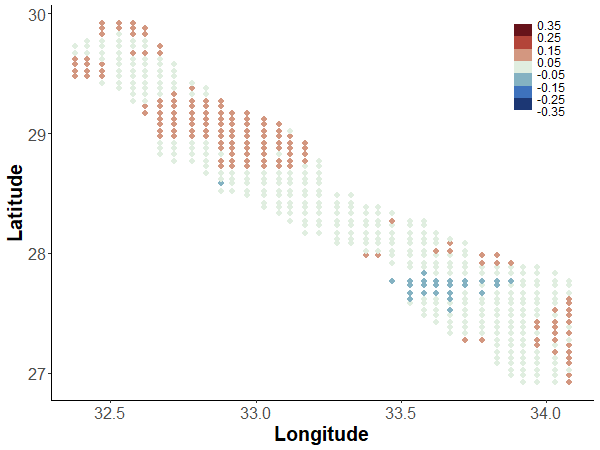}
  \caption{Differences in $\Tilde{\chi}_{0.95}(\boldsymbol{s}_k)$ between periods $(1985$-$1989)-(2011$-$2015)$ for all spatial locations $\boldsymbol{s}_k$, $k\in\{1,\ldots,D\}$. [Left to right and top to bottom:] A-B-C-D margins.}
  \label{fig:chi diff}
\end{figure}

\begin{figure}[t]
  \flushleft
  \includegraphics[width=0.48\textwidth]{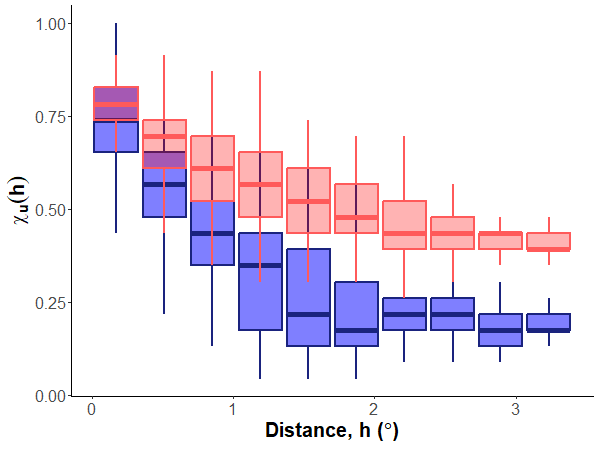}
  \hfill
  \includegraphics[width=0.48\textwidth]{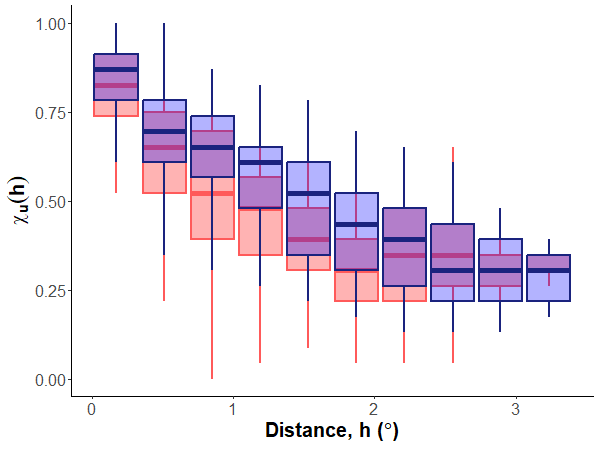}
  \hfill
  \includegraphics[width=0.48\textwidth]{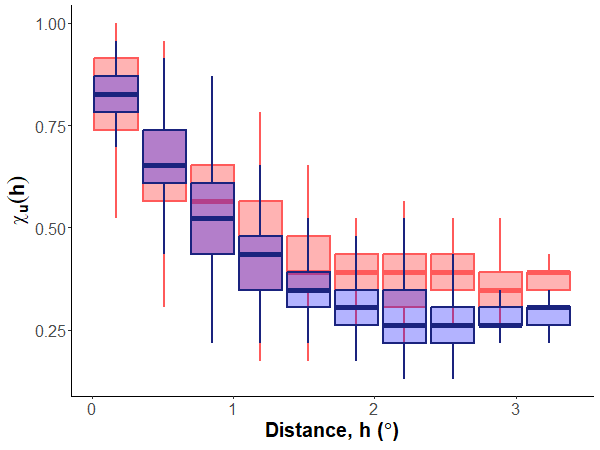}
  \hfill
  \includegraphics[width=0.48\textwidth]{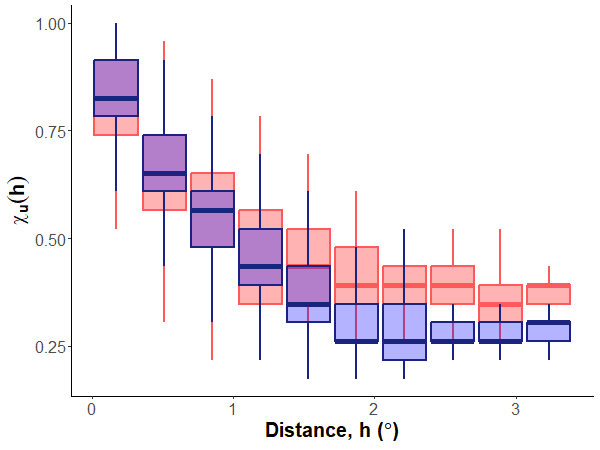}
  \caption{Boxplots of $\chi_u(h_{jk})$ estimates grouped in $10$ equidistant distance blocks. Boxplots in red are based on $\chi_u(h_{jk})$ values for the period $1985-1989$, while boxplots in blue are based on $\chi_u(h_{jk})$ values for years $2011-2015$. All estimates are calculated using $u=0.95$. [Left to right and top to bottom:] A-B-C-D margins.}
  \label{fig:chi cloud}
\end{figure}

Having obtained marginally detrended versions of the dataset, as described in Sections \ref{sec:marg a} -- \ref{sec:marg d}, we perform exploratory investigations of the spatial extremal dependence structure. We focus on two key measures to assess changing spatial extremal dependence over time. The first measure examined is an average version of the $\chi_u$ and $\eta_u$ measures defined in \eqref{spat chi} and \eqref{spat eta} respectively. To be precise, we define $\Tilde{\chi}_u(\boldsymbol{s}_k) = \frac{1}{D-1}\sum_{j\neq k}\chi_u(h_{jk})$ and $\Tilde{\eta}_u(\boldsymbol{s}_k) = \frac{1}{D-1}\sum_{j\neq k}\eta_u(h_{jk})$. Note that $\boldsymbol{s}_k$ is the conditioning site from the definition of $\chi_u(h_{jk})$ in \eqref{spat chi}, and is therefore included as an argument in the definition of $\Tilde{\chi}_u(\boldsymbol{s}_k)$ and $\Tilde{\eta}_u(\boldsymbol{s}_k)$ to highlight that averaging is performed for each conditioning location separately. These quantities allow us to assess average changes in the extremal dependence of the entire spatial domain over time when computed over and contrasted against different time periods. For example, Figure \ref{fig:chi diff} shows differences in $\Tilde{\chi}_u(\boldsymbol{s}_k)$ between periods $1985-1989$ and $2011-2015$, i.e.\ $\Tilde{\chi}_{u,(1985-1989)}(\boldsymbol{s}_k)-\Tilde{\chi}_{u,(2011-2015)}(\boldsymbol{s}_k)$ where $\Tilde{\chi}_{u,(A-B)}(\boldsymbol{s}_k)$ represents the measure $\Tilde{\chi}_{u}(\boldsymbol{s}_k)$ calculated using data in the time period $A-B$. The quantity $\Tilde{\chi}_{u,(1985-1989)}(\boldsymbol{s}_k)-\Tilde{\chi}_{u,(2011-2015)}(\boldsymbol{s}_k)$ is calculated empirically for all spatial locations $\boldsymbol{s}_k$, $k=\{1,\ldots,D\}$, using $u=0.95$. A similar plot is obtained respectively for the quantity $\Tilde{\eta}_{u,(1985-1989)}(\boldsymbol{s}_k)-\Tilde{\eta}_{u,(2011-2015)}(\boldsymbol{s}_k)$ and is available in Figure \ref{fig:eta diff} of Appendix \ref{sec:appendix_eta}. The second measure we look at is pairwise $\chi_u(h_{jk})$ estimates from \eqref{spat chi}. We compute those empirically for the same two periods and threshold level ($u=0.95$) and plot them against distance, thus assessing how the extremal dependence of our dataset changes with respect to distance and over time. These estimates are grouped in ten equidistant distance blocks and visualised via boxplots in Figure \ref{fig:chi cloud} for better clarity of representation.

It is important to note the considerably different messages Figures \ref{fig:chi diff} and \ref{fig:chi cloud} convey concerning the change in dependence over time: when $\Tilde{\chi}_{u,(1985-1989)}-\Tilde{\chi}_{u,(2011-2015)}$ is computed using margins A (top left plot of Figure \ref{fig:chi diff}), a weakening in spatial extremal dependence is suggested, with $\Tilde{\chi}_u$'s being smaller for the later period, $2011-2015$, compared to the earlier period, $1985-1989$, across all available locations. The opposite however is suggested by the plot corresponding to margins B (top right plot of Figure \ref{fig:chi diff}), with a sign reversal in $\Tilde{\chi}_u$ suggesting a strengthening of spatial extremal dependence over time. Finally, plots based on margins C and D (bottom left and right plots of Figure \ref{fig:chi diff} respectively) suggest little to no change over time. The same conclusion holds when examining $\Tilde{\eta}_{u,(1985-1989)}-\Tilde{\eta}_{u,(2011-2015)}$ values (see Figure \ref{fig:eta diff}), or plots of $\chi_u$ over distance shown in Figure \ref{fig:chi cloud}.

Given the non-negligible discrepancies between detrending procedures, we aim to quantify what repercussions these might have on a more practical level by fitting the same spatial model to the four datasets presented in Section \ref{sec:methods}. In doing so we are not attempting to model potential non-stationarity in the spatial dependence, but rather to highlight the different conclusions obtained concerning spatial extremal dependence from the different marginal models. In practice, this feature is not usually considered in extremal analyses of spatial processes. We fit the CSE model of \citet{Wadsworth2022}, introduced in Section \ref{sec:cse} and detailed in Appendix \ref{sec:appendix_cse}. Parameter estimation is achieved via a composite maximum likelihood scheme --- information from all locations acting as the conditioning site was allowed to contribute to the inference procedure --- using code provided as supplementary material from \citet{Wadsworth2022}. The resulting estimates are presented in Table \ref{tab:cse fits}. A constant high threshold corresponding to the $95\%$ quantile of the standard Laplace distribution was used to obtain these results.

\begin{figure}
  \flushleft
  \includegraphics[width=0.48\textwidth]{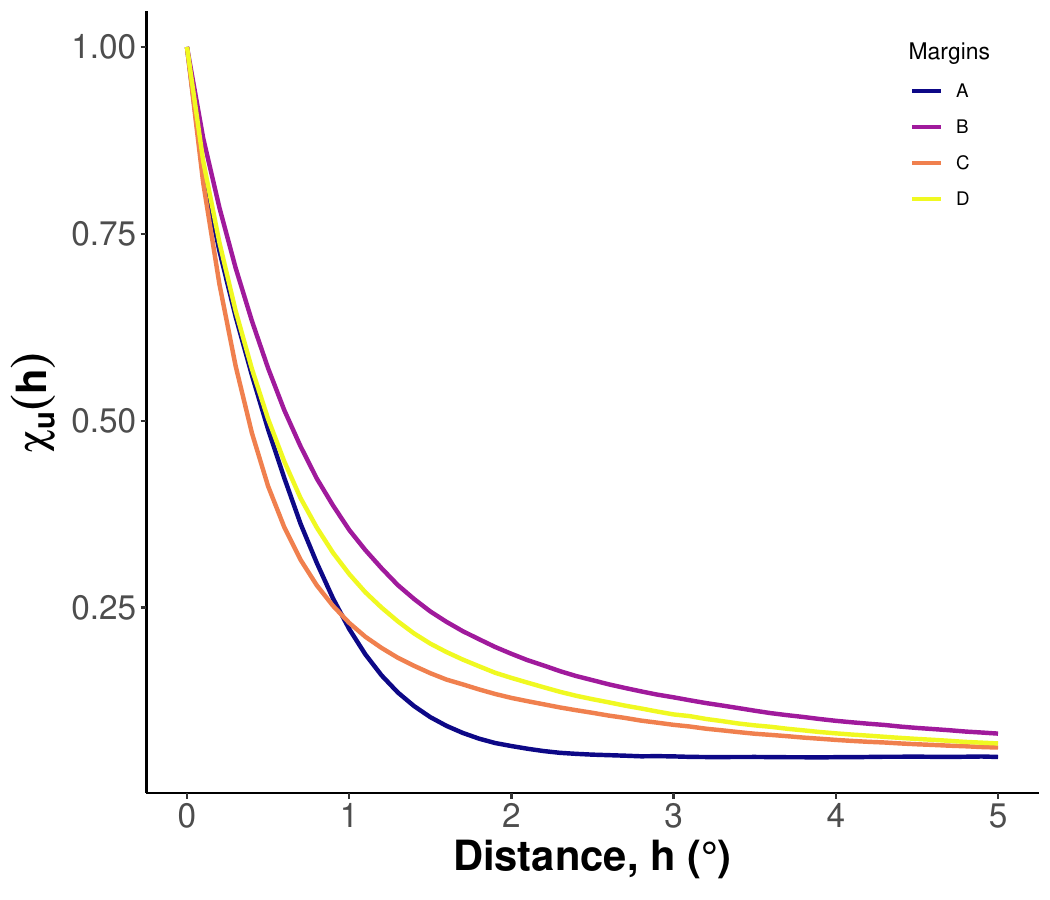}
  \hfill
  \includegraphics[width=0.48\textwidth]{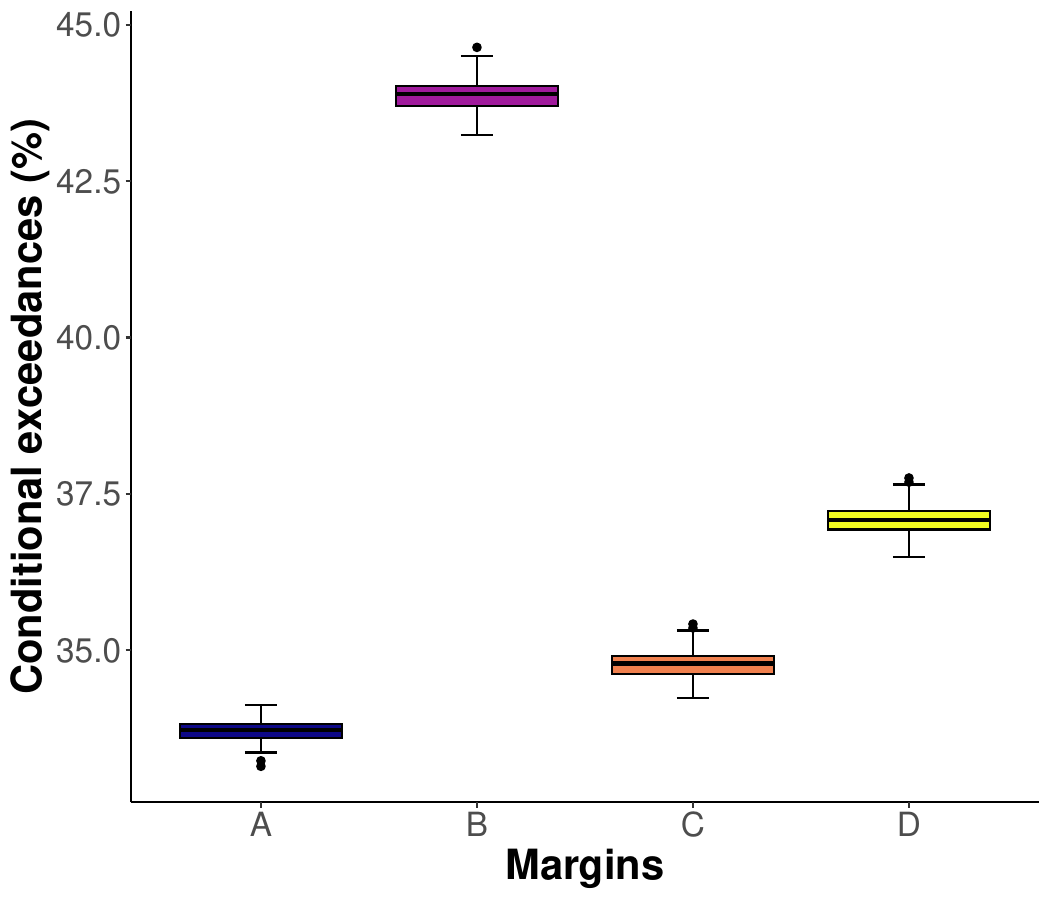}
  \caption{[Left:] Model $\chi_u(h)$ estimates for $u=0.95$ plotted over distance. Lines correspond to mean $\chi_u(h)$ estimates. [Right:] Boxplots of model simulated conditional exceedances. All estimates are obtained based on $200$ simulations of $10^4$ random fields each.}
  \label{fig:model sim}
\end{figure}

\begin{table}[t]
\centering
\begin{tabular}{ccccccccc} 
\toprule
\textbf{Margins} & $\hat{\kappa}$ & $\hat{\lambda}$ & $\hat{\beta}$ & $\hat{\phi}_Z$ & $\hat{\mu}_Z$ & $\hat{\delta}$ & $\hat{\nu}_Z$ & $\hat{\sigma}_Z$ \\ \midrule
A                & $2.00$                   & $1.02$                    & $0.68$                  & $76.1$                & $-8.51$               & $1.03$                   & $1.34$                & $5.21$                   \\
B                & $0.901$                   & $3.23$                    & $1.000$                  & $85.2$                & $-163$             & $1.04$                   & $1.64$                & $7.28$                   \\
C                & $0.975$                     & $1.63$                    & $1.000$                  & $15.4$                 & $-7.81$                & $0.829$                     & $1.64$                  & $1.84$                   \\
D                & $0.968$                   & $2.26$                    & $1.000$                  & $15.9$                & $-11.4$              & $0.982$                   & $1.57$                & $1.80$                   \\ \bottomrule
\end{tabular}
\caption{Table of CSE model coefficients. Parameters $\kappa$ and $\lambda$ relate to the function $a_{\boldsymbol{s}-\boldsymbol{s}_0}(\cdot)$, parameter $\beta$ relates to the function $b_{\boldsymbol{s}-\boldsymbol{s}_0}(\cdot)$, and parameters $\phi_Z$, $\mu_Z$, $\delta$, $\nu_Z$ and $\sigma_Z$ relate to the structure of the process $Z^0$; see Appendix \ref{sec:appendix_cse} as well as  \citet{Wadsworth2022} for more details.}
\label{tab:cse fits}
\end{table}
Combined parameter estimates from the CSE model are not immediately interpretable in terms of the strength of spatial dependence. We therefore assess this by simulating from the fitted models to obtain model-based estimates of $\chi_u(h)$, which we compare graphically in the left plot of Figure \ref{fig:model sim}. Additionally, the right hand plot of Figure \ref{fig:model sim} provides percentages of conditional threshold  exceedances obtained from our model simulations. That is, conditional on a single location $\boldsymbol{s}_0$ (selected to lie approximately in the centre of the spatial domain) being extreme, the percentage of exceedances of the $95\%$ quantile observed in the remaining locations was calculated.  Both plots once again succeed in conveying the disagreement between the same spatial analysis of differently detrended datasets, with the weakest overall dependence for margins A and the strongest for margins B, and highlight the important influence marginal considerations can inflict on dependence features and vice versa. Note that these summaries do not include the effects of parameter uncertainty in the model fits, so may be interpreted with some caution. Uncertainty quantification could potentially be obtained by block bootstrap, but is both computationally intensive and difficult to do appropriately given the likely non-stationary nature of some of the margins. Therefore this is not performed for this investigative analysis.

\section{Discussion} \label{sec:discussion}
The focus of this paper has primarily been on unpicking the margins-dependence interrelationship in the analysis of spatial extremes. By examining four different marginal detrending techniques, each with different merits and drawbacks, and comparing their effects on the dependence characteristics of the correspondingly detrended datasets we have established two key findings. 

Our first message is that margins and dependence are intricately connected. Their interaction affects both apparent non-stationarity in spatial extremal dependence over time, as well as spatial extremal dependence estimates in model fits assuming temporal stationarity. This is highlighted both by visual exploratory tools such as the ones presented in Figures \ref{fig:chi diff}, \ref{fig:eta diff} and \ref{fig:chi cloud} and model fits like the ones implemented in Section \ref{sec:results} and summarised in Table \ref{tab:cse fits} and Figure \ref{fig:model sim}. The high sensitivity of conclusions to the marginal model is something we should be aware of when performing similar analyses. 

Our second finding is that the task of marginal detrending itself can be a very difficult undertaking in practice. One reason is that there is no single approach to marginal modelling. We strive for the most stationary data possible, but it is difficult to guarantee that this has been achieved to an acceptable level, especially with a large number of sites (in this case $>500$) to consider. Another reason is that, when attempting to analyse a dataset in practice, there is usually no way of knowing a priori the level of its underlying complexity and therefore making an informed decision on the best detrending approach to adopt. One has to rely on exploratory --- principally graphical --- tools which might not straightforwardly reveal the existence and extent of non-stationarity in the dataset and can be plagued by other complicating factors unrelated to non-stationarity, such as temporal dependence discussed in Section \ref{sec:intro}.

\section*{Declarations}

\subsection*{Acknowledgments}
We are grateful to the two anonymous reviewers and editors for constructive comments and suggestions that have improved this article. 
This paper is based on work completed while Lydia Kakampakou was part of EPSRC funded project EP/W524438/1. The Red Sea data were provided by GHRSST, Met Office and CMEMS.

\subsection*{Data Availability}
Sea surface temperature data can be downloaded at \url{https://marine.copernicus.eu}.

\subsection*{Conflict of interest}
The authors declare no potential conflict of interests.

\bibliography{references}

\newpage
\appendix

\section*{Appendix}
\renewcommand{\thesubsection}{\Alph{subsection}}

\subsection{Additional information on the CSE model}\label{sec:appendix_cse}
\setcounter{equation}{0}
\renewcommand\theequation{A\arabic{equation}}
Let $\{Y(\boldsymbol{s}):\boldsymbol{s}\in S \subset \mathbb{R}^2\}$ be a stationary and isotropic spatial random field with exponential-tailed (typically standard Laplace) margins. Then, as mentioned in Section \ref{sec:cse} of the main text, the CSE model takes the expression
\begin{equation}\label{cse model apdx}
    \left\{Y(\boldsymbol{s}) \mid Y(\boldsymbol{s}_0)>t: \boldsymbol{s} \in \mathcal{S}\right\} \stackrel{\text{d}}{\approx} \left\{a_{\boldsymbol{s}-\boldsymbol{s}_0}\left(Y(\boldsymbol{s}_0)\right) + b_{\boldsymbol{s}-\boldsymbol{s}_0}\left(Y(\boldsymbol{s}_0)\right)Z^0(\boldsymbol{s}): \boldsymbol{s}\in S\right\},
\end{equation}
where $a_{\boldsymbol{s}-\boldsymbol{s}_0}: \mathbb{R} \rightarrow \mathbb{R}$ and $b_{\boldsymbol{s}-\boldsymbol{s}_0}: \mathbb{R} \rightarrow (0, \infty)$, $\boldsymbol{s} \in S$. 

The authors take $a_{\boldsymbol{s}-\boldsymbol{s}_0}(y)=y\alpha(\boldsymbol{s}-\boldsymbol{s}_0)$, with $a_{0}(y)=y$, allowing $\alpha(\cdot)$ to be a powered-exponential correlation function of distance. That is, 
\begin{equation}\label{alpha fun}
   \alpha(\boldsymbol{s}-\boldsymbol{s}_0) = 
   \begin{cases}
    1, \quad & \|\boldsymbol{s}-\boldsymbol{s}_0\|<\Delta, \\
    \exp \left\{-\left(\|\boldsymbol{s}-\boldsymbol{s}_0\|-\Delta\right)^\kappa/\lambda\right\}, \quad & \|\boldsymbol{s}-\boldsymbol{s}_0\|>\Delta, 
   \end{cases} 
\end{equation}
where $\kappa\in (0,2]$, $\lambda>0$ and $\Delta\geq0$. The above specification allows for modelling of asymptotic dependence up to a distance $\Delta$ from the conditioning site and asymptotic independence beyond that point, with exponentially decreasing strength of dependence. As for the function $b_{\boldsymbol{s}-\boldsymbol{s}_0}$ the authors propose three different parametric forms, each achieving different modelling requirements. We adopt the form $b_{\boldsymbol{s}-\boldsymbol{s}_0}(x) = 1+a_{\boldsymbol{s}-\boldsymbol{s}_0}(x)^\beta$, $\beta>0$, which is advocated by the authors as the most flexible of the three options recommended. This choice is suitable for AI data (i.e.\ $\Delta=0$ in \eqref{alpha fun}) but also allows for the possibility that the process $\{Y(\boldsymbol{s})\}$ is independent of $Y(\boldsymbol{s}_0)$ when $\|\boldsymbol{s}-\boldsymbol{s}_0\|$ is large. For details on this matter we refer the reader to the original paper.

The residual process $\{Z^0(\boldsymbol{s})\}$ must satisfy the constraint $Z^0(\boldsymbol{s}_0)=0$. To ensure this is achieved, we let $\{Z_G(\boldsymbol{s})\}$ be a stationary Gaussian process with mean $\mu_Z$ and covariance function $\text{Cov}(\boldsymbol{s},\boldsymbol{s}+h)=\sigma_Z^2 \exp \left\{-\left(h/\phi_Z \right)^{\nu_Z} \right\}$, where $\sigma_Z>0$, $\phi_Z>0$, $\nu_Z>0$, and $h$ is the spatial lag. We then take $\{Z^0(\boldsymbol{s})\}$ to have the same dependence structure as $\{Z_G(\boldsymbol{s})\mid Z_G(\boldsymbol{s}_0)=0\}$. This modelling choice is also provided by the authors. 
Finally, the delta-Laplace distribution is used for the marginal modelling of $\{Z^0(\boldsymbol{s})\}$, which includes both Gaussian ($\delta=2$) and Laplace ($\delta=1$) distributions as special cases and is more flexible than either owing to the ``transition" parameter $\delta$. The density of the delta-Laplace distribution is
\begin{equation*}
    f_{\delta L}(z)=\frac{\delta}{2\sigma_{\delta L}\Gamma(1/\delta)}\exp\left\{-\left|(z-\mu_{\delta L})/\sigma_{\delta L}\right|^{\delta}\right\}, \quad \delta>0, \mu_{\delta L} \in \mathbb{R}, \sigma_{\delta L}>0.
\end{equation*}
For a more detailed account on the CSE model we refer the reader to the original \citet{Wadsworth2022} paper.

\newpage
\subsection{Supplementary figures for Section 3.4}\label{sec:appendix_margD}
\setcounter{figure}{0}        
\renewcommand\thefigure{B\arabic{figure}}  
Figure \ref{fig:histD} displays a histogram of the automatically selected marginal thresholds for margins D, obtained by implementing the methodology of \citet{murphy2023automated}.
\begin{figure}[h]
  \centering
  \includegraphics[width=0.4\textwidth]{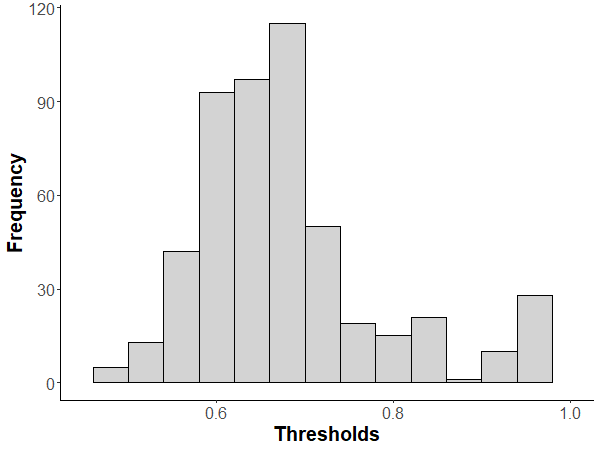}
  \caption{Histogram of automatically selected thresholds used in margins D.}
  \label{fig:histD}
\end{figure}

\subsection{Supplementary figures for Section 4}\label{sec:appendix_eta}
\renewcommand\thefigure{C\arabic{figure}}  
Figure \ref{fig:eta diff} shows differences in $\Tilde{\eta}_u(\boldsymbol{s}_k)$ between periods $1985-1989$ and $2011-2015$, i.e.\ $\Tilde{\eta}_{u,(1985-1989)}(\boldsymbol{s}_k)-\Tilde{\eta}_{u,(2011-2015)}(\boldsymbol{s}_k)$, where $\Tilde{\eta}_{u,(A-B)}(\boldsymbol{s}_k)$ represents the measure $\Tilde{\eta}_{u}(\boldsymbol{s}_k)$ calculated using data in the time period $A-B$. The quantity $\Tilde{\eta}_{u,(1985-1989)}(\boldsymbol{s}_k)-\Tilde{\eta}_{u,(2011-2015)}(\boldsymbol{s}_k)$ is calculated empirically for all spatial locations $\boldsymbol{s}_k$, $k=\{1,\ldots,D\}$, using $u=0.95$. 
\begin{figure}[h]
  \flushleft
  \includegraphics[width=0.48\textwidth]{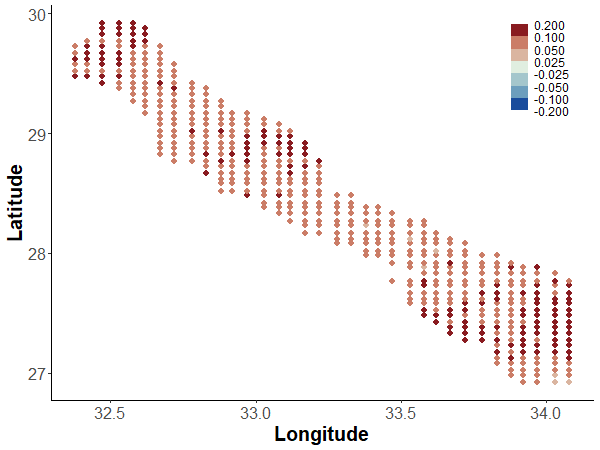}
  \hfill
  \includegraphics[width=0.48\textwidth]{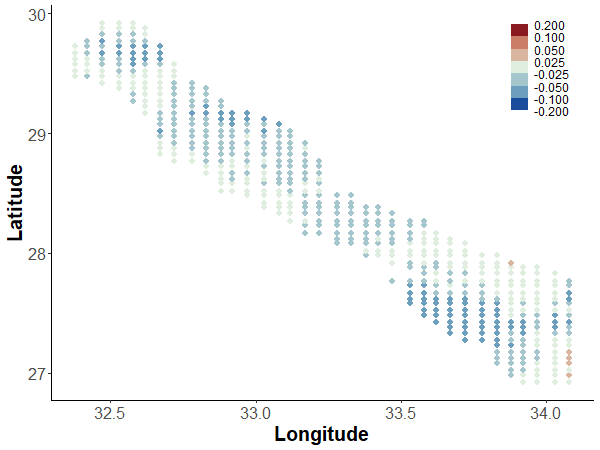}
  \hfill
  \includegraphics[width=0.48\textwidth]{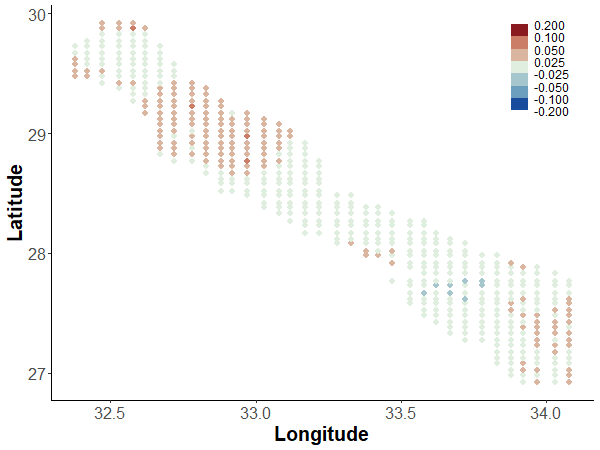}
  \hfill
  \includegraphics[width=0.48\textwidth]{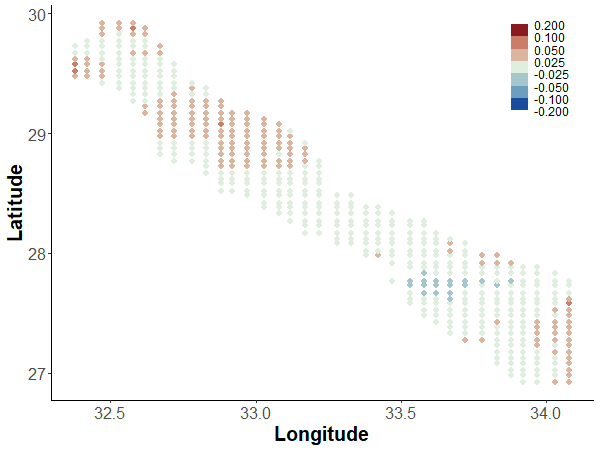}
  \caption{Differences in $\Tilde{\eta}_{0.95}(\boldsymbol{s}_k)$ between periods $(1985-1989) - (2011-2015)$ for all spatial locations $\boldsymbol{s}_k$, $k\in\{1,\ldots,D\}$. [Left to right and top to bottom:] A-B-C-D margins.}
  \label{fig:eta diff}
\end{figure}

\end{document}